\documentclass[aps,prl,twocolumn,showpacs,groupedaddress]{revtex4}

\usepackage{amssymb}   

\usepackage{graphicx}
\usepackage{dcolumn}
\usepackage{bm}
\usepackage{amssymb}   

\begin{document}

\title{Critical Phenomena and the Quantum Critical Point of Ferromagnetic Zr$_{1-x}$Nb$_{x}$Zn$_{2}$}

\author{D. A. Sokolov$^{1}$, M. C. Aronson$^{1}$, W. Gannon$^{1}$, Z. Fisk$^{2}$}

\address{$^{1)}$ University of Michigan, Ann Arbor, MI 48109-1120, USA\\
$^{2)}$ Florida State University, Tallahassee, FL 32310-3706, USA\\}

\begin{abstract}
{We present a study of the magnetic properties of
Zr$_{1-x}$Nb$_{x}$Zn$_{2}$, using an Arrott plot analysis of the
magnetization. The Curie temperature T$_{C}$ is suppressed to zero
temperature for Nb concentration x$_{C}$=0.083$\pm$0.002, while
the spontaneous moment vanishes linearly with T$_{C}$ as predicted
by the Stoner theory. The initial susceptibility $\chi$ displays
critical behavior for x$\leq$ x$_{C}$, with a critical exponent
which smoothly crosses over from the mean field to the quantum
critical value. For high temperatures and x$\leq$x$_{C}$, and for
low temperatures and x$\geq$x$_{C}$ we find that
$\chi^{-1}$=$\chi_{o}^{-1}$+aT$^{4/3}$, where $\chi_{o}^{-1}$
vanishes as x$\rightarrow$x$_{C}$. The resulting magnetic phase
diagram shows that the quantum critical behavior extends over the
widest range of temperatures for x=x$_{C}$, and demonstrates how a
finite transition temperature ferromagnet is transformed into a
paramagnet, via a quantum critical point.}
\end{abstract}
 \maketitle


Zero temperature phase transitions and their attendant quantum
critical fluctuations have emerged as dominant features in the
phase diagrams of different types of strongly correlated electron
systems, from oxide superconductors~\cite{orenstein2000} and heavy
fermion compounds~\cite{doniach1977,stewart2001}, to low
dimensional materials~\cite{emery1979}. These fluctuations
qualitatively modify the electronic states near a quantum critical
point (QCP), leading to unusual temperature divergences of the
susceptibility and heat capacity, to anomalous power law behavior
in the electrical transport, and even to scale invariance in the
magnetic responses
~\cite{montfrooij2003,aronson2001,aronson1995,schroder2000,
hayden1991}. The fundamental excitations near QCPs are
qualitatively unlike those of conventional metals, representing in
some cases entirely new classes of collective
states~\cite{laughlin2000, senthil2004}. A central issue is
whether these unusual properties require the exceptionally rich
physics of these host systems, derived from low dimensionality and
strong correlations, or whether only proximity to a zero
temperature phase transition is required. Thus, it is important to
identify electronically simple systems, and to study the evolution
of their critical phenomena as the ordered phases are suppressed
to zero temperature.

Itinerant ferromagnets are particularly attractive hosts for such
a study, as they lack the complex interplay of itinerant and
localized character found near the QCPs of heavy fermion
systems~\cite{schroder2000, si2001}. Pressure and compositional
variation have been used to suppress the finite temperature
magnetic ordering transition, finding that the magnetically
ordered phase can vanish discontinuously as in pressurized
MnSi~\cite{pfleiderer1997}, UGe$_{2}$~\cite{pfleiderer2002}, and
perhaps
ZrZn$_{2}$~\cite{uhlarz2004,knapp1971,smith1971,huber1975,grosche1995},
or continuously as in (Ni$_{1-x}$Pd$_{x}$)$_{3}$Al
~\cite{sato1975}. While disorder can affect the order of the
quantum critical phase transition in itinerant
ferromagnets~\cite{belitz2002}, it is generally found in systems
with continuous transitions tuned by a parameter $\Gamma$ that the
QCP which occurs for $\Gamma$=$\Gamma_{c}$ and T=0 dominates the
magnetic phase diagram and generates a phase line
T$_{C}^{(d+n)/z}$$\sim$($\Gamma$-$\Gamma_{c}$), where d=3,
z=n+2=3~\cite{millis1993,pfleiderer1997,grosche1995,julian1996}.
Near the QCP, the electronic part of the heat capacity is
maximized \cite{thompson1989, wada1990,fuller1993,tateiwa2001,
vollmer2002} while the electrical resistivity evolves from
$\rho\propto$T$^{1+\delta}$ for $\Gamma=\Gamma_{C}$ to the Fermi
liquid $\rho\propto$T$^{2}$ for $\Gamma
>\Gamma_{C}$\cite{oomi1998,pfleiderer2001,
steiner2003,grosche1995}.  The low field magnetization is
anomalous near the QCP \cite{grosche1995,pfleiderer1997}, but a
detailed study spanning the ordered and paramagnetic phases
remains lacking. We provide this study of the magnetization of
ZrZn$_{2}$ here, discussing how the QCP is generated with Nb
doping, and the subsequent evolution of the critical phenomena as
the QCP is approached.

Zr$_{1-x}$Nb$_{x}$Zn$_{2}$ is ideal for such a study. Neutron form
factor measurements~\cite{pickart1964} show that the magnetic
moment is spatially delocalized, consistent with the small
spontaneous moment~\cite{matthias1958}. We establish a magnetic
phase diagram, and show that it is dominated by a QCP at x$_{C}$ =
0.083$\pm$0.002 and T$_{C}$ = 0 K. Stoner theory describes the
ferromagnetism of ZrZn$_{2}$ well, indicating that variation in
the density of states at the Fermi level controls both the Curie
temperature T$_{C}$ and the zero temperature spontaneous moment
m$_{0}$(0). Our measurements of the initial magnetic
susceptibility $\chi$(T) describe how the critical phenomena
evolve with Nb doping, crossing over from mean field behavior when
the reduced temperature is low and when Nb concentrations are far
from the critical value, to a regime at small x-x$_{C}$ where the
susceptibility is controlled by the QCP over an increasingly broad
range of temperatures.


Polycrystalline Zr$_{1-x}$Nb$_{x}$Zn$_{2}$ samples with Nb
concentrations $0\leq x\leq 0.14$ were prepared by solid state
reaction ~\cite{ogawa1967}. X-ray diffraction confirmed the C-15
ZrZn$_{2}$ structure~\cite{matthias1958} at each composition, as
well as residual amounts of unreacted Zr and Zn. The magnetization
was measured using a Quantum Design magnetometer at temperatures
from 1.8 K to 200 K and in magnetic fields up to 7 T. The inset of
Fig.~1a shows the magnetic isotherms for Zr$_{1-x}$Nb$_{x}$Zn$_{2}$
(x = 0.03) presented as an Arrott plot. Both the spontaneous
magnetization m$_{0}$(T) = $\lim_{H\rightarrow 0}$ m(H, T) and the
initial susceptibility $\chi$(T) = $\lim_{H \rightarrow 0}$ dm(H,
T)/dH were determined by extrapolation of data from fields larger
than 4.5 T. Previous de Haas-van Alphen experiments on a single
crystal of ZrZn$_{2}$ found a field induced transition in the
magnetization near 5 T ~\cite{kimura2004}, and a complex pressure -
temperature phase diagram was proposed for this
sample~\cite{uhlarz2004}. However, our  Arrott plots are linear in
fields from 1 T - 7 T, at least at low temperatures and for small x,
indicating that this field-driven transition is absent, as it was in
earlier work ~\cite{ogawa1967,huber1975,seeger1995}. The temperature
dependence of m$_{0}$(T) is plotted in Fig.~1a, showing that Nb
doping continuously reduces T$_{C}$ and the zero temperature
spontaneous moment m$_{0}$(0). Fig.~1a shows that for each
x$\leq$x$_{C}$, m$_{0}$ is described by the mean field expression
m$_{0}$($\tau$) = m$_{0}$(0)$\tau^{\beta}$, where $\beta$ = 0.5 and
$\tau$ = (T$_{C}$-T)/T$_{C}$. The suppression of T$_{C}$ with Nb
doping is shown in Fig.~1b, demonstrating that the ferromagnetic
phase line obeys the expected
T$_{C}^{(d+n)/z}$=T$^{4/3}$$\propto$(x-x$_{C}$) (d=z=3,
n=1)~\cite{millis1993}, terminating at a critical concentration
x$_{C}$=0.083$\pm$0.002, analogous to the results of high pressure
measurements~\cite{grosche1995}.

\begin{figure}
\includegraphics[scale=0.65]{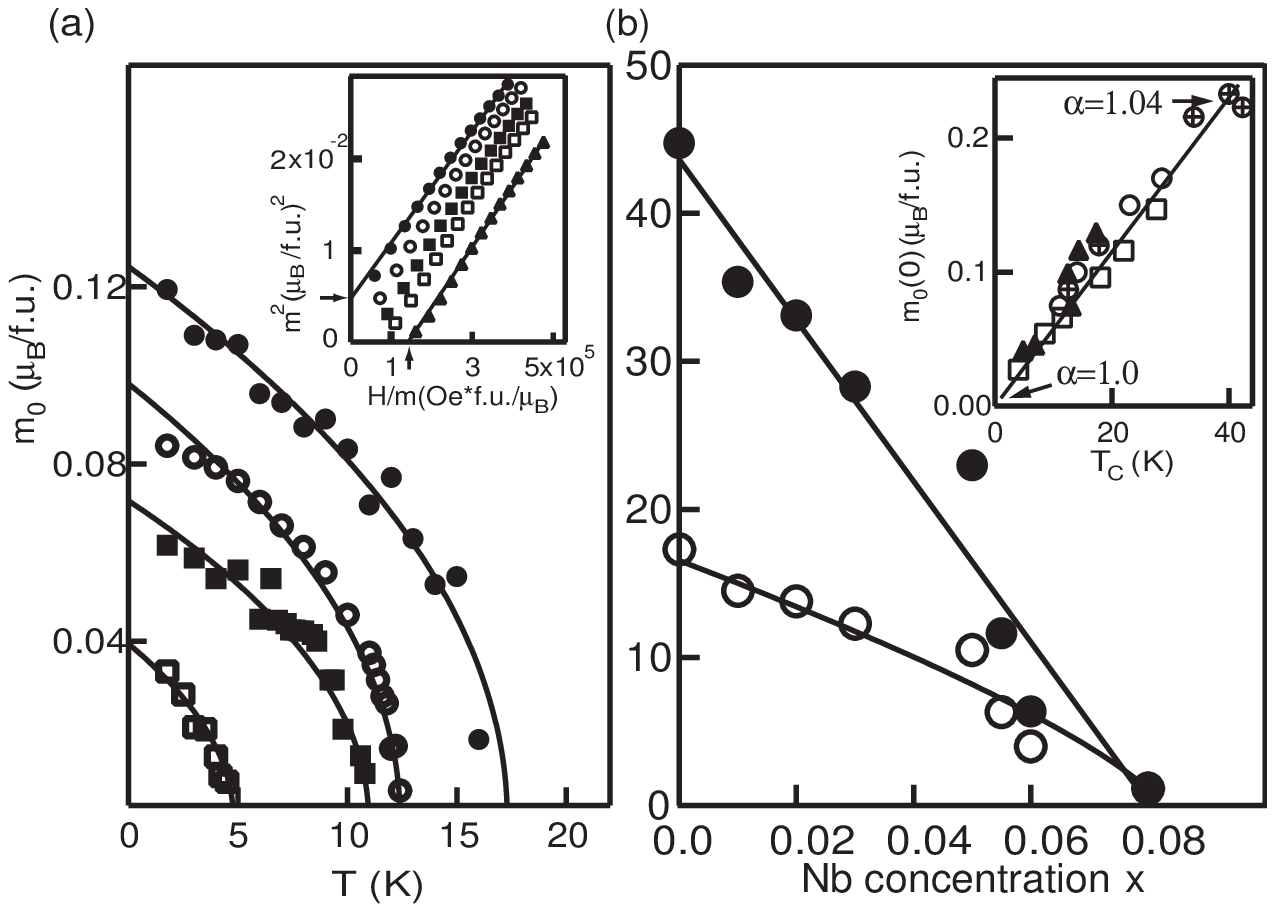}
\caption{\label{fig:epsart} Temperature dependence of the
spontaneous magnetization m$_{0}$ in $\mu_{B}$ per formula
unit(f.u.) for Zr$_{1-x}$Nb$_{x}$Zn$_{2}$ (x=0 ($\bullet$), 0.03
($\circ$), 0.05 ($\blacksquare$), 0.06 ($\square$)). The solid lines
are fits to m$_{0}$(T) = m$_{0}$(0)[(T$_{C}$-T)/T$_{C}$]$^{0.5}$.
Inset: Example of an Arrott plot for x=0.03, for temperatures from
1.8 K ($\bullet$) to 20 K ($\blacktriangle$), the solid lines are
fits to m$^{2}$$\propto$H/m, arrows mark extrapolated 1/$\chi$(T)
(horizontal axis) and m$^{2}$(T) (vertical axis). (b) Variation of
Curie temperature T$_{C}$ ($\circ$) and T$_{C}^{4/3}$ ($\bullet$)
with Nb concentration x. Solid lines are fits to T$_{C}$$\propto$
(x$_{C}$-x)$^{3/4}$ and T$_{C}^{4/3}$$\propto$(x$_{C}$-x), with
x$_{C}$ = 0.083$\pm$0.002. Inset: The zero temperature spontaneous
moment m$_{0}$(0)$\propto$T$_{C}$. Y,Ti,Hf doping ($\oplus$
)~\cite{ogawa1968}; pressure, single crystal
($\circ$)~\cite{uhlarz2004}, pressure, ZrZn$_{1.9}$ polycrystal
($\square$)~\cite{huber1975}, Nb doping, polycrystals, this work
($\blacktriangle$). Solid line is Stoner parameter $\alpha$
~\cite{knapp1970}.}
\end{figure}

The suppression of T$_{C}$ and m$_{0}$ with Nb doping indicates that
the Stoner theory adequately describes the ferromagnetism in
ZrZn$_{2}$, where the underlying control parameter $\alpha$ is the
product of the Coulomb interaction and the density of states at the
Fermi level. Stoner theory predicts that T$_{c}$ and m$_{0}$(0) are
proportional for $\alpha\sim$1. The inset of Fig.~1b shows this
proportionality is valid not only for our Nb doped samples, but also
for those with Ti, Hf, and Y doping~\cite{ogawa1968}, those with
modified stoichiometry~\cite{knapp1970,knapp1971} and even when high
pressures are applied~\cite{huber1975,grosche1995,uhlarz2004}.
Values of $\alpha$ are indicated in the inset of Fig.~1b, suggesting
that modifications to the underlying electronic structure and not
the disorder associated with doping are primarily responsible for
altering the stability of ferromagnetism in ZrZn$_{2}$. Indeed, only
a two percent reduction in $\alpha$ is necessary to drive the
ferromagnetism in undoped ZrZn$_{2}$ to the brink of instability,
whether by doping or by the application of pressure.

The initial magnetic susceptibility $\chi$ is considerably modified
as the system is driven from a finite temperature instability in
undoped ZrZn$_{2}$, through a QCP, and into the paramagnetic phase.
Arrott plot analyses are used to deduce $\chi^{-1}$(T) = $\lim_{H
\rightarrow 0}$ dH/dm(H, T) shown in Fig.~2 for a wide range of Nb
concentrations x. $\chi$ diverges at T$_{C}$ in the ferromagnetic
samples (x$\leq$x$_{C}$), with little sign of critical rounding. The
sample with x=0.08 displays a nearly power-law response in absolute
temperature, as expected near a QCP
~\cite{millis1993,lonzarich1997}. Finally, for x$\geq$x$_{C}$,
$\chi$(T) approaches a constant value $\chi_{o}$ as T$\rightarrow$0,
signalling that long range ferromagnetic order is no longer
possible.

\begin{figure}
\includegraphics[scale=0.44]{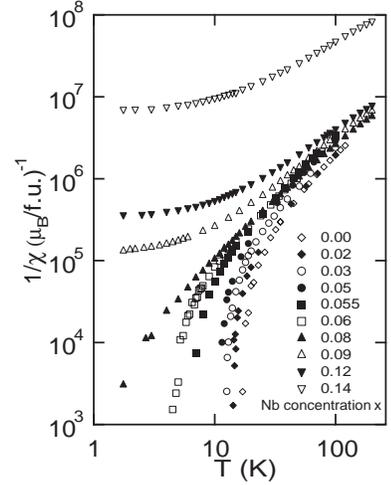}
\caption{\label{fig:epsart} The temperature dependence of the
inverse of the initial susceptibility $\chi^{-1}$ for Nb
concentrations both larger and smaller than the critical
concentration x$_{C}$=0.083$\pm$0.002.}
\end{figure}

The initial susceptibility for x$<$x$_{C}$ is well described by a
power law $\chi$=$\chi_{oo}\tau^{-\gamma}$ over at least two decades
of reduced temperature $\tau$=(T-T$_{C}$)/T$_{C}$, and for absolute
temperatures as large as 100 K (Fig.~3a). The inset of Fig.~3a shows
that $\gamma$ increases smoothly from the near-mean field value
1.08$\pm$0.05 previously observed in ZrZn$_{2}$~\cite{seeger1995},
to 1.33$\pm$0.01 for x$\sim$x$_{C}$. Since the interactions which
lead to magnetic order in itinerant ferromagnets are long ranged,
the intrinsic exponents related to the underlying symmetries are
only found at reduced temperatures which are much smaller than those
accessed in our experiments.
~\cite{seeger1995a,seeger1989,kaul1985}. We conclude that the
variation of $\gamma$ with Nb concentration is the result of a
gradual crossover from the mean field behavior associated with a
finite temperature ferromagnetic transition for x$<$x$_{C}$ to
quantum criticality as x$\rightarrow$x$_{C}$.

\begin{figure}
\includegraphics[scale=0.65]{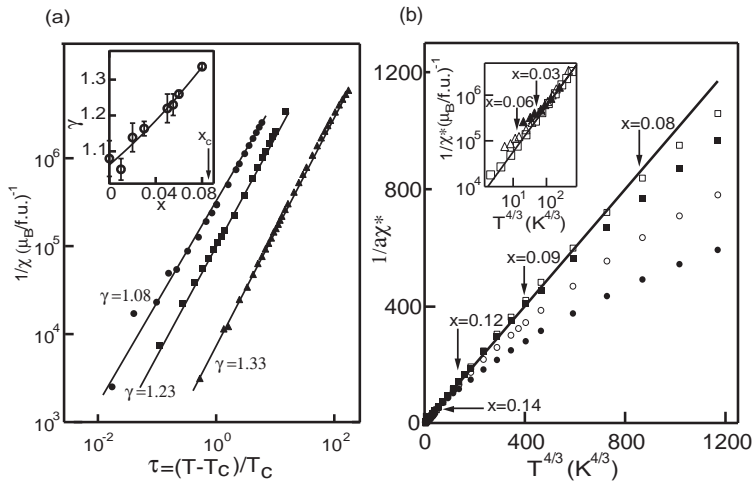}
\caption{\label{fig:epsart} $\chi^{-1}$ =
$\chi_{oo}^{-1}\tau^{\gamma}$ at different Nb concentrations: x=0
($\bullet$), x=0.055 ($\blacksquare$), x=0.08 ($\blacktriangle$).
Solid lines are the best fits to $\chi^{-1}$ =
$\chi_{oo}^{-1}\tau^{\gamma}$. Inset: the critical exponent $\gamma$
($\circ$) increases continuously as x$\rightarrow$x$_{C}$. Solid
line is guide for the eye. (b) 1/a$\chi^{*}$ (defined in the text)
for x$\geq$x$_{c}$ is proportional to T$^{4/3}$ for T$\leq$T$^{*}$.
For x$\geq$x$_{c}$, 1/a$\chi^{*}$ deviates from T$^{4/3}$ (solid
line) $\it{above}$ a cutoff temperature T$^{\ast}$ marked by arrows,
x=0.08 ($\square$), x=0.09 ($\blacksquare$), x=0.12 ($\circ$),
x=0.14 ($\bullet$). Inset: For x$\leq$x$_{c}$ 1/$\chi^{*}$ deviates
from T$^{4/3}$ behavior $\it{below}$ a crossover temperature T$_{G}$
marked by arrows: x=0.08 ($\square$), x=0.06 ($\triangle$), x=0.03
($\blacktriangle$). Solid line is best fit of 1/$\chi^{\ast}$=
aT$^{4/3}$ for x=0.08.}
\end{figure}

Isolating the temperatures and Nb concentrations where $\chi$ is
dominated by the QCP is straightforward in the paramagnetic regime
(x$\geq$x$_{C}$). Near the QCP, the initial susceptibility for an
itinerant, three dimensional ferromagnet is given by $\chi^{-1}$ =
$\chi_{o}^{-1}$ + aT$^{4/3}$, with $\chi_{o}^{-1}\propto$
(x-x$_{C}$)~\cite{millis1993}. The variation of $\chi_{o}^{-1}$ with
Nb concentration x is plotted in the lower panel of Fig.~4. As
predicted, $\chi_{o}^{-1}$ vanishes approximately linearly with
(x-x$_{C}$), while a changes by less than 10 $\%$. The temperature
dependent part of the initial susceptibility is isolated by defining
1/$\chi^{\ast}$=($\chi^{-1}$-$\chi_{o}^{-1}$). 1/a$\chi^{\ast}$ is
plotted in Fig.~3b as a function of T$^{4/3}$ for each of the three
paramagnetic concentrations with x=0.09, 0.12, and 0.14, and for
comparison x=0.08, which has T$_{C}$ = 1.2$\pm$0.l K. Fig.~3b
demonstrates that 1/a$\chi^{\ast}$=T$^{4/3}$ below a temperature
T$^{\ast}$ which vanishes at x=0.15 (Fig.~4), the termination of
quantum criticality for x$>$x$_{C}$.

Quantum criticality is also observed for x$\leq$x$_{C}$, although
the critical phenomena associated with the finite temperature
ferromagnetic transition ultimately dominate as
T$\rightarrow$T$_{C}$. Since $\chi^{-1}$ =
$\chi_{o}^{-1}$+aT$^{4/3}$, with $\chi_{o}^{-1}\propto$(x-x$_{C}$),
the quantum critical susceptibility is largest for x$\sim$x$_{C}$,
and extends over the broadest range of absolute temperatures, since
T$_{C}\rightarrow$0. This is demonstrated in the inset of Fig.~3b,
where we have plotted 1/$\chi^{\ast}$=$\chi^{-1}$-$\chi_{o}^{-1}$ as
a function of T$^{4/3}$. For x$\leq$x$_{C}$, the T$^{4/3}$ behavior
is only found above a temperature T$_{G}$ which grows rapidly with
(x-x$_{C}$), as shown in Fig.~4. The lower panel of Fig.~4 shows
that for x$\leq$ x$_{C}$, 1/$\chi_{o}$ decreases approximately
linearly with (x-x$_{C}$), while the magnitude a of the T$^{4/3}$
term is approximately independent of x, as in the paramagnetic phase
x$>$x$_{C}$.

\begin{figure}
\includegraphics[scale=0.4]{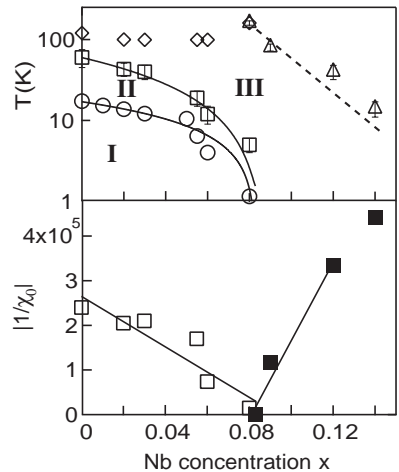}
\caption{\label{fig:epsart} Top: The temperature - Nb concentration
phase diagram for Zr$_{1-x}$Nb$_{x}$Zn$_{2}$. T$_{C}$(x) ($\circ$),
with the phase line T$_{C}\propto$(x-x$_{C}$)$^{3/4}$, the cutoff
temperature T$_{G}$ ($\square$) for x$\leq$x$_{C}$ with the
crossover line T$_{G}\propto$(x-x$_{C}$)$^{1.2\pm0.3}$, the cutoff
temperature T$^{\ast}$ ($\triangle$) for x$\geq$x$_{C}$, dashed line
is a guide for the eye. ($\diamond$) indicate the highest
temperature investigated for x$\leq$x$_{C}$. I, II, and III
correspond, respectively, to the region of stable ferromagnetism,
the crossover region, and the quantum critical region as defined in
the text. Bottom: Variation with Nb concentration of the T=0 inverse
susceptibility 1/$\chi_{o}$ for x$\geq$x$_{C}$ ($\blacksquare$)and
for x$\leq$x$_{C}$ ($\square$). Solid lines are fits with
1/$\chi_{o}$ $\propto$ $|$x-x$_{C}$$|$. For x=0.14, 1/$\chi_{o}$ is
divided by 15 to appear on the plot.}
\end{figure}

Our magnetization measurements establish that the phase diagram of
Zr$_{1-x}$Nb$_{x}$Zn$_{2}$ has three different regimes, depicted in
Fig.~4. Ferromagnetic order is found in Region I, below a phase line
T$_{C}$(x) which is controlled by the x=x$_{C}$, T$_{C}$=0 QCP. The
Stoner criterion which describes the stability of ferromagnetism
remains unchanged even as x$\rightarrow$x$_{C}$, indicating that the
reduction in the electronic density of states drives the QCP. The
spontaneous moment obeys the mean field expression,
m$_{o}$($\tau$)=m$_{o}$(0)$\tau^{-0.5}$ in Region I.

Simple power law divergences with reduced temperature $\tau$ are
found in Region II, $\chi$($\tau$)=$\chi_{o}\tau^{-\gamma}$, and the
enhancement of $\gamma$ as x$\rightarrow$ x$_{C}$ reveals that
Region II is best considered a crossover region, controlled by the
relative magnitudes of $\tau$ and (x-x$_{C}$). Specifically, for
small $\tau$ and large (x-x$_{C}$), we find the mean field behavior
of a finite transition temperature ferromagnet $\chi\sim \tau^{-1}$.
In the opposite limit (small (x-x$_{C}$) and large $\tau\sim$T) we
find that the QCP is dominant, yielding $\chi^{-1} \sim $
a+bT$^{4/3}$. Accordingly, it is possible to identify the quantum
critical behavior of a three dimensional ferromagnet
$\chi^{-1}$=$\chi_{o}^{-1}$+aT$^{4/3}$ above a temperature T$_{G}$
which decreases rapidly as x$\rightarrow$x$_{C}$.

The quantum critical regime III is also extensive for paramagnetic
concentrations x$>$x$_{C}$, occurring below a temperature
T$^{\ast}$ which is almost 150 K for x=x$_{C}$, and dropping
rapidly at larger x. Region III extends to the highest
temperatures investigated for x$<$x$_{C}$. We speculate that the
boundary of the quantum critical regime for x$\geq$x$_{C}$
coincides with the condition that the correlation length is
reduced to some minimal length, such as the lattice constant. The
phase diagram suggests that this occurs at T=0 for x$\sim$ 0.15,
but at increasingly high temperatures T$^{\ast}$ as x approaches
x$_{C}$ from above.

The quantum critical behavior documented in this work is in
excellent agreement with theoretical predictions for a three
dimensional T$_{C}$=0 ferromagnet. The T$_{C}$=0, x=x$_{C}$ QCP
affects a surprisingly broad area of the x-T phase diagram,
competing with the conventional critical phenomena for even
x$<<$x$_{C}$. We find no indication of new collective phases near
the QCP in Zr$_{1-x}$Nb$_{x}$Zn$_{2}$, and suggest that further
measurements at lower temperatures and with refined samples would be
very interesting to further pursue this issue.

Work at the University of Michigan was performed under grant
NSF-DMR-0405961. Work at FSU was supported by NSF-DMR-0433560. We
acknowledge useful conversations with A. J. Millis, M. B. Maple, K.
V. Kamenev, M. E. J. Newman and L. Sander, as well as the
hospitality and support of J. E. Crow and the National High Magnetic
Field Laboratory in Tallahassee during the early stages of this
work.


\begin{thebibliography}{20}
\bibitem{orenstein2000}J. Orenstein and A. J. Millis, Science
$\bf{288}$, 468 (2000).

\bibitem{doniach1977}S. Doniach, Physica $\bf{91B}$, 231 (1977).

\bibitem{stewart2001}G. R. Stewart, Rev. Mod. Phys. $\bf{73}$, 797
(2001).

\bibitem{emery1979}V. J. Emery in \it{Highly Conducting
One-Dimensional Solids}\rm, ed. J. T. Devreese, R. P. Evrard, and V.
E. van Doren (New York, Plenum, 1979) p. 247.

\bibitem{montfrooij2003}W. Montfrooij, M. C. Aronson
 $\textit{et al}$., Phys. Rev. Lett. $\bf{91}$, 087202 (2003).

\bibitem{aronson2001} M. C. Aronson $\textit{et al}$., Phys. Rev. Lett. $\bf{87}$, 197205 (2001).

\bibitem{aronson1995}M. C. Aronson $\textit{et al}$., Phys. Rev. Lett.
$\bf{75}$, 725 (1995).

\bibitem{schroder2000}A. Schr\"{o}der $\textit{et al}$., Nature (London) $\bf{407}$, 351
(2000).

\bibitem{hayden1991}S. M. Hayden $\textit{et al}$., Phys. Rev. Lett. $\bf{66}$, 821 (1991).

\bibitem{laughlin2000}R. B. Laughlin $\textit{et al}$., Proc. Nat. Acad. Sci. $\bf{97}$, 32
(2000).

\bibitem{senthil2004}T. Senthil $\textit{et al}$., Science $\bf{303}$, 1490 (2004).

\bibitem{si2001}Q. Si $\textit{et al}$., Nature $\bf{413}$, 804 (2001).

\bibitem{pfleiderer1997}C. Pfleiderer $\textit{et al}$.,
Phys. Rev. B. $\bf{55}$, 8330 (1997).

\bibitem{pfleiderer2002}C. Pfleiderer and A. D. Huxley, Phys. Rev.
Lett. $\bf{89}$, 147005 (2002).

\bibitem{uhlarz2004} M. Uhlarz $\textit{et al}$.,
Phys. Rev. Lett. $\bf{93}$, 256404 (2004). It has subsequently
been shown that samples of this type are nonstoichiometric,
especially near their surfaces (E. A. Yelland $\textit{et al}$.,
cond-mat/0502341). The elevated Curie temperature of these samples
may also be a signature of Zn deficiency~\cite{knapp1971}. The
pressure driven QCP in near-stoichiometric ZrZn$_{2}$ was
previously shown to be second order~\cite{smith1971, huber1975,
grosche1995}.

\bibitem{knapp1971}G. S. Knapp $\textit{et al}$., J. Appl. Phys. $\bf{42}$,
1341 (1971).

\bibitem{smith1971}T. F. Smith $\textit{et al}$., Phys. Rev.
Lett. $\bf{27}$, 1732 (1971).

\bibitem{huber1975}J. G. Huber $\textit{et al}$., Solid State Commun. $\bf{16}$, 211 (1975).

\bibitem{grosche1995}F. M. Grosche $\textit{et al}$., Physica B $\bf{206-207}$, 20 (1995).

\bibitem{sato1975}M. Sato, J. Phys. Soc. Japan $\bf{39}$,
98 (1975).

\bibitem{belitz2002}D. Belitz, T. R. Kirkpatrick,
Phys. Rev. Lett. $\bf{89}$, 247202 (2002).

\bibitem{millis1993}A. J. Millis, Phys. Rev. B $\bf{48}$, 7183 (1993).

\bibitem{julian1996}S. R. Julian $\textit{et al}$., J. Phys.: Condensed Matter $\bf{8}$, 9675 (1996).

\bibitem{thompson1989}J. D. Thompson $\textit{et al}$, Physica B $\bf{161}$, 317 (1989).

\bibitem{wada1990}H. Wada $\textit{et al}$, J. Phys. Soc. Japan $\bf{59}$, 2956
(1990).

\bibitem{fuller1993}C. J. Fuller $\textit{et al}$, J. Appl. Phys $\bf{73}$, 5338 (1993).

\bibitem{tateiwa2001}N. Tateiwa $\textit{et al}$, J. Phys.: Condensed Matter
$\bf{13}$, L17 (2001).

\bibitem{vollmer2002}R. Vollmer $\textit{et al}$, Physica B $\bf{312-313}$, 112
(2002).

\bibitem{oomi1998}G. Oomi $\textit{et al}$, J. Alloys
and Compounds $\bf{271-273}$, 482 (1998).

\bibitem{pfleiderer2001}C. Pfleiderer $\textit{et al}$, Nature  $\bf{414}$, 427 (2001).

\bibitem{steiner2003}M. J. Steiner $\textit{et al}$, Physica B $\bf{329-333}$, 1079 (2003).

\bibitem{pickart1964}S. J. Pickart $\textit{et al}$., Phys. Rev. Lett. $\bf{12}$, 444 (1964).

\bibitem{matthias1958}B. T. Matthias and R. M. Bozorth, Phys. Rev. $\bf{1091}$,
604 (1958).

\bibitem{ogawa1967}S. Ogawa and N. Sakamoto, J. Phys. Soc. Japan $\bf{22}$, 1214 (1967).

\bibitem{kimura2004}N. Kimura $\textit{et al}$. Phys. Rev. Lett. $\bf{92}$, 197002 (2004).

\bibitem{seeger1995}M. Seeger $\textit{et al}$., J. Magn. Magn. Mater. $\bf{139}$, 312 (1995).

\bibitem{ogawa1968}Shinji Ogawa, J. Phys. Soc. Japan $\bf{25}$, 109 (1968).

\bibitem{knapp1970}G. S. Knapp, J. Appl. Phys. $\bf{41}$, 1073 (1970).

\bibitem{lonzarich1997}G. G. Lonzarich, in $\it {Electron}$ edited by
M. Springford (Cambridge University Press, Cambridge, England,
1997).

\bibitem{seeger1989}M. Seeger and H. Kronmuller, J. Magn. and Magn. Mater. $\bf{78}$,
393 (1989).

\bibitem{kaul1985}S. N. Kaul, J. Magn. and Magn. Mater. $\bf{53}$,
5 (1985).

\bibitem{seeger1995a}M. Seeger $\textit{et al}$., Phys. Rev. B. $\bf{51}$, 12585 (1995).

\end{thebibliography}
\end{document}